\documentclass[preprint,prc]{revtex4}
\usepackage{epsfig}
\usepackage{amssymb}

\newcommand{\dbar}{d \hspace*{-1.25ex}\raisebox{.7ex}{-}}
\begin{document}
\bibliographystyle{apsrev}

\title{ Microscopic calculation of the spin-dependent neutron scattering lengths
on $^{\text{3}}$He }
\author{ H.~M.~Hofmann }
\affiliation{Institut f{\"u}r Theoretische Physik III,
   University of Erlangen-N{\"u}rnberg, Staudtstra\ss{}e 7,
  D 91058 Erlangen, Germany}
\author{ G.~M.~Hale }
\affiliation{Theoretical Division, Los Alamos National Laboratory,
  Los Alamos, NM 87545, USA}

\begin{abstract}
We report on the spin-dependent neutron scattering length on $^{\text{3}}$He 
from  a microscopic calculation of $p-^3$H, $n-^3$He, and $d-^2$H
  scattering employing the Argonne $v_{18}$  nucleon-nucleon
  potential with and without additional three-nucleon force. The results and
  that of a comprehensive $R$-matrix analysis are compared to a recent
  measurement. The overall agreement for the scattering lengths is
  quite good. The imaginary parts of the scattering lengths are very sensitive
  to the inclusion of three-nucleon forces, whereas the real parts are
  almost insensitive.
\end{abstract}
\maketitle

\section*{Introduction}

The scattering length is an easy way to compare low energy scattering data
with calculations. Recently the spin-dependent scattering lengths for neutrons
on tritons were calculated \cite{ANT} with the correlated hyperspherical
harmonics technique. The calculations displayed a weak dependence on the various
nucleon-nucleon (NN) and three-nucleon (NNN) potentials used. Afterwards an
resonating group model (RGM) calculation gave for a modest model space for the
triton \cite{BP} very close results for the Argonne $v_{18}$ (AV18) \cite{AV18}
and AV18 + Urbana IX (UIX) \cite{AV8P-U8-U9} potentials. Neutron scattering on $^3$He
is much more difficult to handle. Since at the n-$^3$He threshold the charge
exchange channel is already open by about 700keV, the neutron absorption
cross section is much higher than the elastic one for low energies. Therefore the
scattering lengths become complex. The imaginary parts are rather well
determined from the experiments by \cite{Als-nie}.
The real parts of the spin-dependent neutron scattering lengths of $^3$He were
recently measured \cite{Zimmer} with much higher precision than before
\cite{Kohl}. These new results could only be compared to rather old theoretical
approaches. Almost 30 years ago Kharchenko and Lebashew \cite{KL} calculated the
real parts of the scattering lengths $a_0$ = 7.52 fm and $a_1$ = 3.07 fm,
neglecting the Coulomb force and using a simple separable S-wave potential.
Sears and Khanna \cite{KH} gave a Breit-Wigner estimate of the same values.

We organize the paper in the following way: The next section contains
a brief discussion of the Resonating Group Model (RGM) calculation and
the model spaces used. Then we compare $R$-matrix and RGM results of
the neutron scattering length for various interactions with the data
and discuss the effect of NNN-forces.

\section{RGM and model space}

We use the Resonating Group Model \cite{RRGM, RRGM-VIEWEG, RRGM-TANG}
to compute the scattering in the $\rm ^4$He  system
using the Kohn-Hulth\'en variational principle \cite{KOHN}. The main
technical problem is the evaluation of the many-body matrix elements
in coordinate space. The restriction to a Gaussian basis for the
radial dependencies of the wave function allows for a fast and
efficient calculation of the individual matrix elements \cite{RRGM,
  RRGM-TANG}. However, to use these techniques the potentials must
also be given in terms of Gaussians. In this work we use suitably
parametrized versions of the AV18 \cite{AV18} 
$NN$ potential and the UIX \cite{AV8P-U8-U9}
and $V_3^*$ proposed in \cite{V3-SCHADOW} and used in \cite{BP} $NNN$
potentials.

In the $^4$He system we use a model space with six two-fragment
channels, namely the $p - ^3$H, the $n - ^3$He, the $d - ^2$H,
the $d - ^2$H(S=0), the \dbar resonance, the \dbar - \dbar and the
$(pp) - (nn)$ channels. The last three are an approximation to
the three- and four-body breakup channels that cannot in practice
be treated within the RGM. The $\rm ^4$He is treated as four clusters
in the framework of the RGM to allow for the required internal orbital
angular momenta of $\rm ^3$H, $\rm ^3$He or $\rm ^2$H.

For the scattering calculation we include all $S$, $P$ and 
$D$ wave contributions to the $J^\pi = 0^+, 1^+, 2^+, 0^-, 1^- \text{ and }
2^-$ channels. From the $R$-matrix analysis these channels are known
to reproduce the low-energy experimental data.  The full wave function for these
channels contains over 100 different spin and orbital angular momentum
configurations, hence it is too complicated to be given in detail. 
We started with the 29 - dimensional model space for $^3$H/$^3$He as described
in \cite{BP}, increased it to dimension 35 by adding components to the
wave function with two $D$ - waves on the internal coordinates of the triton,
optimized for AV18 and UIX together.
By this modest increase of the model space, we gained 650 keV binding energy.
Since this change in model space resulted in noticable effects on observables
\cite{HLRB} we aimed at an almost converged model space. Using a genetic
algorithm \cite{CWGEN} for AV18 and UIX together allowing for $S$, $P$
and $D$ waves on all internal coordinates we found a triton binding energy of
-8.460 MeV for dimension 70.
This result compares favourably with the numerically exact one of Nogga 
\cite{NOGGA-FAD} of -8.478 MeV. Since the Gaussian width parameters were
optimized for $NN$ and $NNN$ - interaction together, the agreement for the
AV18 alone is only -7.57 MeV, compared to the exact one of -7.62 MeV.
For the deuteron we used 5 width parameters for the S-wave and 3 for the
D-wave, yielding -2.213 MeV, just 10 keV short of the experimental value.
The binding energies and relative thresholds for the various potentials are
given in table \ref{thres}. For $NN$ and $NNN$ together the experimental
binding energies and thresholds are very well reproduced.

This representation of $^3$H/$^3$He , deuteron and the unbound $NN$ systems
form the model space of the $^4$He scattering system. We get for the
different $J^{\pi}$ values 5 to 10 physical channels, insufficient to find reasonable
results. 
So-called distortion or pseudo-inelastic
channels \cite{RRGM-TANG} without an asymptotic part
have to be added to improve the description
of the wave function within the interaction region. 
For this purpose all the configurations calculated for the physical channels
but one per channel can be reused, keeping only those width parameters
which describe the internal region.
 Recently  Fonseca \cite{NT-FONSECA} pointed out that states
 having a negative parity $J_3^-$ in the three-nucleon fragment
 increase the $n -^3$H cross section noteably. Contrary to the neutron-triton
 system we found in the $^4$He system in the preliminary small model space
 calculations that the inclusion of such distortion states gave minor effects
 compared to adding UIX. Therefore in the converged calculation we did
 not allow for such states, in order to save computational resources, as we had 
 anyhow to deal with sometimes more than a thousand channels. 

\begin{table}
\centering
 \caption{\label{thres} Comparison of experimental and
 calculated total binding energies and relative thresholds (in MeV) for
 the various potential models used}
 \vskip 0.2cm
 \begin{tabular}{c|c|c|c|c}
 \hline\noalign{\smallskip}
 potential & \multicolumn{2}{c}{$E_{bin}$} & \multicolumn{2}{c}{$E_{thres}$} \\
     & $^3$H & $^3$He & $^3{\rm He}-p$  &  $ { d - d} $ \\
     \noalign{\smallskip}\hline\noalign{\smallskip}
     AV18    & -7.572& -6.857 & 0.715 & 3.145 \\
     AV18 + UIX & -8.460 & -7.713 & 0.747 & 4.033 \\
     AV18 + UIX +$V_3^*$ & -8.452 & -7.705 & 0.747 & 4.025 \\
     exp.    & -8.481 & -7.718 & 0.763 & 4.033\\
     \noalign{\smallskip}\hline
     \end{tabular}
     \end{table}

\section{R-matrix analysis}

The charge-independent R-matrix analysis of the $^4$He system from which
the $n+^3$He scattering lengths are obtained in this paper is similar to
the one described in Section 3 of our previous publication \cite{HE4}.
The isospin-1 R-matrix parameters were determined separately from an
analysis of p+$^3$He scattering data, checked by limited comparisons
with $n+^3$H data, and used essentially fixed in the analysis of the
$^4$He system data in which the isospin-0 parameters were allowed to
vary. New data have been added in most of the reactions, but those
relevant for determining the $n+^3$He scattering lengths include the
neutron total cross sections of refs. \cite{Als-nie, NT-TOTAL-59, HAE83,
KEI03}, the elastic scattering cross sections of \cite{ALF81}, and the
$t(p,n)$ reaction cross-section measurements of \cite{GIB59, BRU99}. 
Charge independence relates the reduced-width amplitudes in the $p+t$ and
$n+^3$He channels, imposing additional constraints on the neutron
scattering lengths from the $p+t$ scattering data at proton energies near
1 MeV.

\section{Determination of the scattering length}

The standard approach to the  scattering length starts from the partial wave
expansion of the scattering amplitude
\begin{equation}
f(\Theta) = \frac{1}{ 2 i k} \sum_{\ell = 0}^{\infty} (2 \ell +1 )(\exp ( 2 i \delta_{\ell}) - 1) P_{\ell}( \cos \Theta)
\end{equation}
with $k$ the neutron wave-vector. For thermal neutrons only the S-wave survives
so that 
\begin{equation}
f(\Theta) = \frac{1}{ 2 i k} (\exp ( 2 i \delta_0) - 1) =\frac{1}{ 2 i k} (S_0  - 1) = (k \cot \delta_0 - i k)^{-1} \label{s0}
\end{equation}
Since $\delta_0$ is an odd function of $k$, one can expand
\begin{equation}
k \cot \delta_0 = -1/a + 1/2 r_e k^2 + O(k^4) \label{expco}
\end{equation}
in which $a$ is the scattering length and $r_e$ the effective range parameter.
Suppressing from now on the subscript on $\delta_0$, we obtain in
lowest order that $a = - \tan \delta / k$.

For neutron - $^3$He scattering the proton - triton channel is already
open, with a large neutron absorption cross section; hence,  $a$ has to
be complex. For thermal neutron scattering ($ k a \ll 1$), the total
scattering and absorption cross section are given (for example in ref.
\cite{SEARS}) by  $\sigma_s = 4 \pi | a |^2$ and
$\sigma_a = 4 \pi a''/ k$ with $a''$ the negative imaginary part of $a$.
Unfortunately, due to numerical limitations, the RGM approach cannot be
used at energies low enough that terms of order $ka$ can be neglected,
and the expressions above need to be modified. Neglecting the effective
range and higher order contributions in eq.~(\ref{expco}), we can write
eq.~(\ref{s0}) as

\begin{equation}
(S - 1)/(2 i k) = ( -1/a  - i k)^{-1}.    \label{t}
\end{equation}

Solving for $ a$ we find
$a =  (1 - S)/(1 + S)/(  i k)$
from which the real and imaginary part can be easily evaluated as
\begin{equation}
\Re a = -\frac{| S | \sin( 2 \delta)}{ 2 k} \frac{1}{4(1 - |S |)^2 + | S | \cos^2 \delta}
\end{equation}
and 
\begin{equation}
\Im a = \frac{(| S |^2 - 1)/ k}{4(1 - |S |)^2 + | S | \cos^2 \delta}
\end{equation}

Note that for $| S |$ = 1 the above expressions go to $- \tan \delta /k$ and 
zero, respectively, as they should. In the case of the R-matrix
analysis, however, the scattering lengths are obtained directly from the
zero-energy limit of eq.~(\ref{t}),
\begin{equation}
a=\lim_{k\rightarrow 0}{1-S\over 2ik}.
\end{equation}

\begin{figure}[h]
\begin{center}
\epsfxsize=10cm
\epsfbox{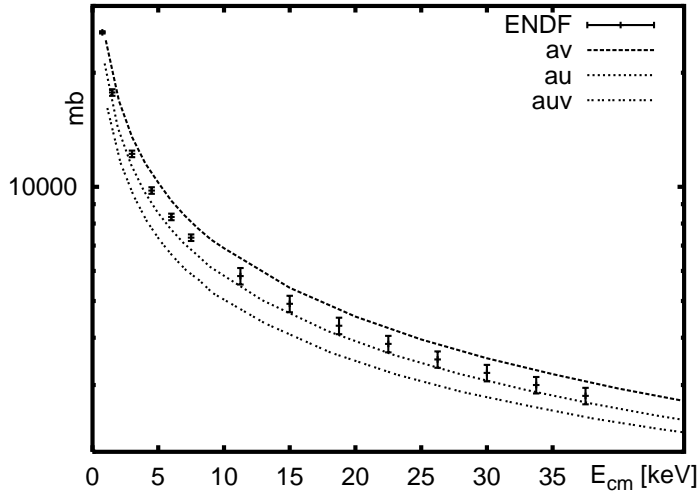}
\caption{\label{stan}
Comparison of the standard neutron cross section of $^3$He
(crosses) \cite{ENDF} and various calculations, AV18 alone (av-conv), AV18 + Urbana IX
(au-conv), and AV18 + Urbana IX + $V_3^*$ \cite{V3-SCHADOW} (auv-conv).
}
\end{center}
\end{figure}

Before we discuss the calculation of the scattering length in the actual case,
let us first compare the standard neutron cross section $^3$He(n,p){$^3$}H.
In fig. \ref{stan} the evaluated standard cross section \cite{ENDF} is
compared to various calculations. 
This standard total neutron cross section is a bit over-predicted by the AV18
NN-force alone, a bit on the lower side for AV18 + UIX and severly under-predicted
by AV18 + Urbana IX + $V_3^*$, see fig. \ref{stan}. At the lowest energy
calculated the S-matrix elements are used to determine the scattering lengths
according to eq. (5) and (6).

For n - $^3$He scattering S-waves occur
as singlet in $J^{\pi} = 0^+$ and as triplet in the $J^{\pi} = 1^+$ channels.
At low neutron energies the $0^+$ channel is dominated by the well known
resonance \cite{TILLEY-A4}, leading to a strong coupling between the two
charge conjugate channels. At 5 keV neutron energy this coupling 
S-matrix element
is about 0.5, the triplet and the P-wave $0^-$ ones are about a factor 20
smaller, all others at least another order of magnitude smaller,
except for the $^1{\rm P}_1$ matrix element,which is only another factor of 4 
smaller. Therefore
we can for every $J^{\pi}$ consider only a two-channel S-matrix. We calculated
according to eqs. (5) and (6) the real and imaginary parts of $a_0$ and 
$a_1$ for various potentials. In table \ref{scatt} the results are compared
to data and the results of the R-matrix analysis.

\begin{table}
\centering
 \caption{\label{scatt} Comparison of experimental and
 calculated real and imaginary scattering lengths (in fm) for
 the various potential models used}
 \vskip 0.2cm
 \begin{tabular}{c|c|c|c|c}
 \hline\noalign{\smallskip}
 potential & \multicolumn{2}{c}{$a_{0}$} & \multicolumn{2}{c}{$a_{1}$} \\
     & $\Re$ & $\Im $ & $\Re$  &  $ \Im  $ \\
     \noalign{\smallskip}\hline\noalign{\smallskip}
     AV18    & 7.81(2)& -4.96(2) & 3.468(1) & -0.0067(1) \\
     AV18 + UIX & 7.62(1) & -4.07(3) & 3.333(1) & -0.0052(1) \\
     AV18 + UIX +$V_3^*$ & 7.57(5) & -3.42(1) & 3.310(1) & -0.0049(1) \\
     R-matrix & 7.400(3) & -4.449(1) & 3.281(2)& -0.0013(2)\\
     exp.    & 7.370(58) \cite{Zimmer}& -4.448(5) \cite{Kohl} & 3.278(53)
\cite{Zimmer}& -0.001(2) \cite{Kohl}\\
     \noalign{\smallskip}\hline
     \end{tabular}
     \end{table}

The numbers in brackets on the calculated values indicate the uncertainty
of the scattering lengths, due to higher order effects. 
We calculate $a$ for a center-of-mass
energy  $E_0$ of a few keV. Since for this energy $k a $ is of the order of a few
percent, higher order contributions might yield changes in $ a $ also
of this order.
By calculating $a$ also at $E_0$ +5 keV and $E_0$ +10 keV we estimate the
uncertainty.

The real parts of $a$ agree within 5 percent with the data and are rather insensitive
to the changes in the potentials. The imaginary parts are very sensitive to
these changes, so they are means to learn about the $NNN$-forces. 
Comparing the table of scattering lengths with the standard neutron cross section
we find the clear relation, that the imaginary parts of $a_0$ just mirror the
ratio of calculated versus evaluated cross sections.  When the curve in 
fig. \ref{stan} is above the data, also the scattering length is larger than
the data and vice versa.
Since we are
always dealing with effective two channel systems, unitarity relates
$\Im a$ to the coupling matrix element squared, hence, the $NNN$-interaction
reduces this coupling appreciably. Since the 3N-bound states are rather dense,
a possible conclusion might be that the
short range repulsion in UIX is too strong and also the longer range attraction,
which can be seen in the $^1S_0$ proton-triton phase shifts being too
attractive for AV18 + UIX, see ref. \cite{HLRB}, thus indicating that the
radial dependence of attraction and repulsion should be changed.
\begin{acknowledgements}
  This work is supported by 
  the BMBF (contract 06ER926) and used
  resources at several computer centers (RRZE Erlangen, 
  SSC Karlsruhe and HLRB M\"unchen).  We want to thank G.\ Wellein and
  G.\ Hager at the RRZE for their help.  The U.~S. Department of Energy
  supported the work of G.~M.~H. on this study.
\end{acknowledgements}

\bibliography{pAper}

\end{document}